\documentclass[prb,amsfonts,twocolumn,showpacs,amssymb,floatfix,bibnotes]{revtex4}
\usepackage{bm}
\usepackage{tabularx}
\usepackage{ascmac}
\usepackage{amsmath}
\usepackage{wrapfig}
\usepackage{graphicx}
\usepackage{float}
\usepackage[FIGBOTCAP]{subfigure}
\usepackage{color}

\bmdefine{\bolds}{s}
\bmdefine{\boldi}{i}
\bmdefine{\boldj}{j}
\bmdefine{\boldtau}{\tau}
\bmdefine{\boldsigma}{\sigma}
\bmdefine{\boldlambda}{\lambda}
\bmdefine{\boldx}{x}
\bmdefine{\boldX}{X}
\bmdefine{\boldk}{k}
\bmdefine{\boldK}{K}
\bmdefine{\boldq}{q}
\bmdefine{\boldQ}{Q}
\bmdefine{\boldr}{r}
\bmdefine{\boldj}{j}

\begin{document}


\title{
Orbital-cooperative spin fluctuation and orbital-dependent transport 
in ruthenates
}


\author{Naoya Arakawa}
\email{arakawa@hosi.phys.s.u-tokyo.ac.jp} 
\affiliation{Department of Physics, 
The University of Tokyo,
Tokyo 113-0033, Japan}


\date{\today}

\begin{abstract}
Unusual transport properties deviating from the Fermi liquid 
are observed in ruthenates 
near a magnetic quantum-critical point (QCP). 
To understand the electronic properties of the ruthenates 
near and away from an antiferromagnetic (AF) QCP, 
I study the electronic structure and magnetic and transport properties 
for the $t_{2g}$-orbital Hubbard model on a square lattice 
in fluctuation-exchange approximation 
including Maki-Thompson (MT) current vertex correction (CVC). 
The results away from the AF QCP reproduce 
several experimental results of Sr$_{2}$RuO$_{4}$ qualitatively 
and provide new mechanisms about the enhancement of spin fluctuation 
at $\boldQ_{\textrm{IC-AF}}\approx (0.66\pi,0.66\pi)$, 
larger mass enhancement of the $d_{xy}$ orbital than that of the $d_{xz/yz}$ orbital, 
and nonmonotonic temperature dependence of the Hall coefficient. 
Also, the results near the AF QCP 
explain the $T$-linear inplane resistivity in Sr$_{2}$Ru$_{0.075}$Ti$_{0.025}$O$_{4}$ 
and give an experimental test on the obtained temperature dependence of the Hall coefficient. 
I reveal 
spatial correlation including the self-energy of electrons beyond mean-field approximations 
is essential to determine the electronic properties of the ruthenates. 
I also show several ubiquitous transport properties near an AF QCP 
and characteristic transport properties of a multiorbital system 
by comparison with results of a single-orbital system near an AF QCP. 
  
\end{abstract}


\maketitle

Many-body effects cause unusual transport properties 
deviating from the Fermi liquid (FL)~\cite{Moriya-review}. 
For example, 
the $T$-linear inplane resistivity, $\rho_{ab}$, and 
Curie-Weiss-like $T$ dependence of the Hall coefficient, $R_{\textrm{H}}$, 
are observed in a quasi-2D single-orbital system 
near an antiferromagnetic (AF) quantum-critical point (QCP)~\cite{cuprate-exp}. 
Also, 
unusual transport properties are observed in ruthenates (i.e., Ru oxides), 
quasi-2D $t_{2g}$-orbital systems: 
Sr$_{2}$Ru$_{0.075}$Ti$_{0.025}$O$_{4}$, located near an AF QCP, 
shows the $T$-linear $\rho_{ab}$~\cite{Ti214-nFL};  
Ca$_{2-x}$Sr$_{x}$RuO$_{4}$ around $x=0.5$, located near a ferromagnetic QCP, 
shows the $T^{1.4}$ dependence of $\rho_{ab}$ and 
Curie-Weiss-like $T$ dependence of $R_{\textrm{H}}$~\cite{CSRO}. 
Note that Sr$_{2}$RuO$_{4}$ shows the FL behaviors~\cite{resistivity-x2,Hall-x2}. 

The origins of these unusual transport properties of the ruthenates are unclear, 
although its understanding leads to a deeper understanding of 
roles of electron correlation and each orbital in transport properties. 

To clarify these origins, 
we should understand roles of 
electron correlation and each $t_{2g}$ orbital. 
In particular, it is necessary to reveal 
effects of the self-energy of electrons and 
electron-hole four-point vertex function due to electron correlation. 
These will give considerable effects in multiorbital systems 
since these play important roles in the single-orbital Hubbard model on a square lattice 
near an AF QCP~\cite{Kontani-CVC} (referred to as the single-orbital case); 
the characteristic $T$ and $\boldk$ dependence of quasiparticle (QP) damping 
causes the $T$-linear $\rho_{ab}$, and 
the characteristic $T$ and $\boldk$ dependence of Maki-Thompson (MT) 
current vertex correction (CVC) due to MT four-point vertex function~\cite{MT} 
causes the Curie-Weiss-like $T$ dependence of $R_{\textrm{H}}$; 
these characteristic dependence arise from 
the Curie-Weiss-like $T$ dependence of the spin susceptibility at $\boldk=(\pi,\pi)$. 

In this paper, 
I reveal the roles of electron correlation and each $t_{2g}$ orbital 
in several electronic properties of the ruthenates near and away from the AF QCP 
and achieve qualitative agreement with 
experiments~\cite{Ti214-nFL,resistivity-x2,Hall-x2}. 
I show the importance of spatial correlation including 
the self-energy of electrons beyond mean-field approximations (MFAs). 
Also, I show several similarities and differences between the transport properties 
of the present case and the single-orbital case~\cite{Kontani-CVC} 
and propose the emergence of the orbital-dependent transport in other systems. 

To describe the electronic structure of the ruthenates, 
I use the $t_{2g}$-orbital Hubbard model on a square lattice, 
\begin{align}
\hat{H}
&=
\textstyle\sum\limits_{\boldk}
\sum\limits_{a,b=1}^{3}
\sum\limits_{s=\uparrow,\downarrow}
\epsilon_{ab}(\boldk)
\hat{c}^{\dagger}_{\boldk a s} 
\hat{c}_{\boldk b s}+
 U 
\sum\limits_{\boldj}
\sum\limits_{a}
\hat{n}_{\boldj a \uparrow} \hat{n}_{\boldj a \downarrow}\notag\\
&+ U^{\prime}  
\textstyle\sum\limits_{\boldj}
\sum\limits_{a>b}
\hat{n}_{\boldj a} \hat{n}_{\boldj b}
- 
J_{\textrm{H}} 
\sum\limits_{\boldj}
\sum\limits_{a>b}
( 
2 \hat{\bolds}_{\boldj a} \cdot 
\hat{\bolds}_{\boldj b} 
+ 
\frac{1}{2} \hat{n}_{\boldj a} \hat{n}_{\boldj b} 
)\notag\\
&+
J^{\prime} 
\textstyle\sum\limits_{\boldj}
\sum\limits_{a>b}
\hat{c}_{\boldj a \uparrow}^{\dagger} 
\hat{c}_{\boldj a \downarrow}^{\dagger} 
\hat{c}_{\boldj b \downarrow} 
\hat{c}_{\boldj b \uparrow},\label{eq:H}
\end{align}
with 
$\epsilon_{11/22}(\boldk)=
-\frac{\Delta_{t_{2g}}}{3}-2 t_{1} \cos k_{x/y}-2 t_{2} \cos k_{y/x}-\mu$, 
$\epsilon_{12/21}(\boldk)= 4 t^{\prime} \sin k_{x} \sin k_{y}$, 
$\epsilon_{33}(\boldk)=
\frac{2\Delta_{t_{2g}}}{3}-2t_{3}(\cos k_{x}+\cos k_{y})-4t_{4}\cos k_{x} \cos k_{y}-\mu$, 
$\epsilon_{13/23/31/32}(\boldk)=0$, $J^{\prime}=J_{\textrm{H}}$, and $U^{\prime}=U-2J_{\textrm{H}}$. 
Hereafter, 
I label the $d_{xz}$, $d_{yz}$, and $d_{xy}$ orbitals $1$, $2$, and $3$, respectively, 
fix the energy unit at eV, and set $\hbar=c=e=\mu_{\textrm{B}}=k_{\textrm{B}}=1$.

\begin{figure*}[tb]
\includegraphics[width=118mm]{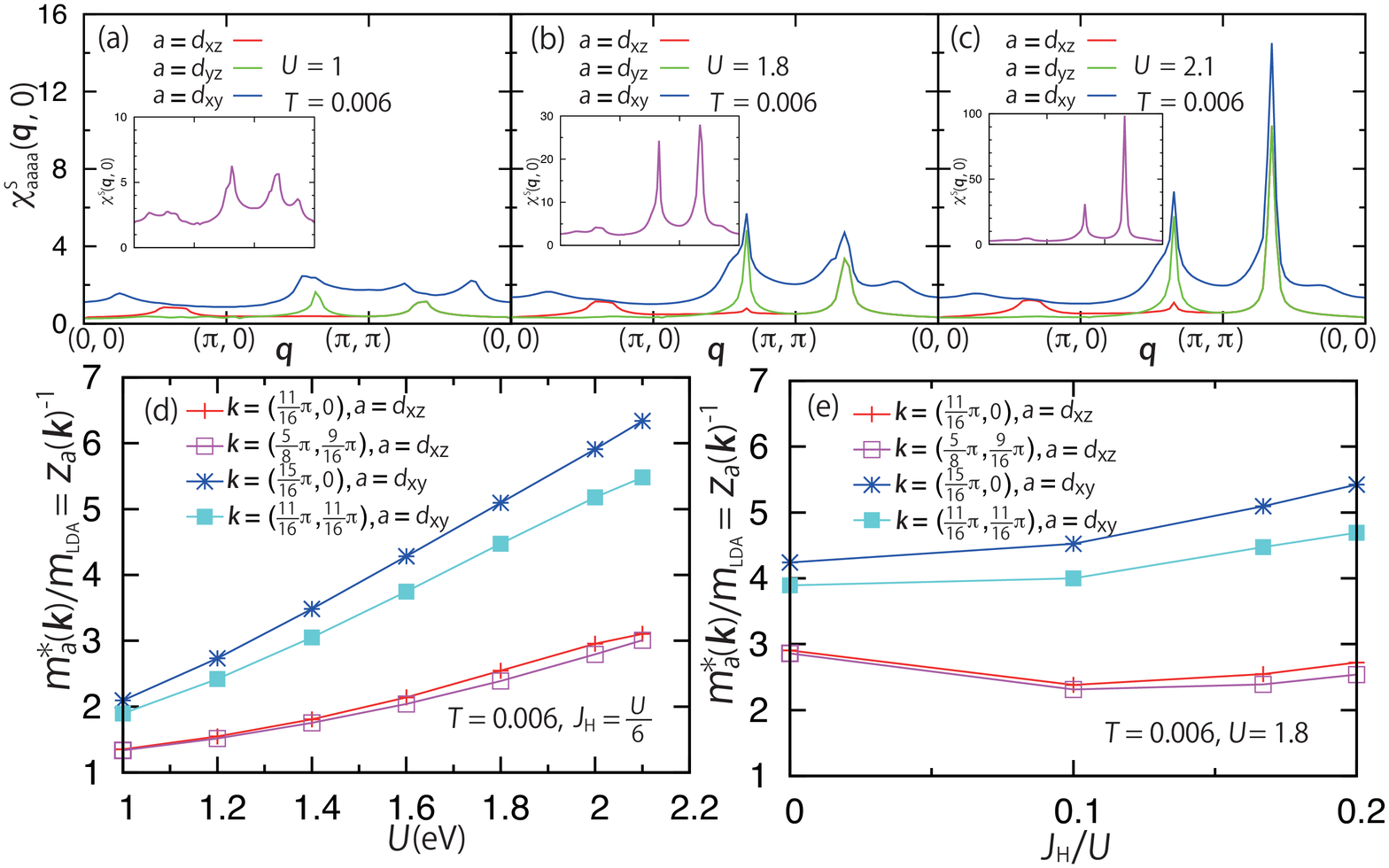}
\vspace{-16pt}
\caption{(Color online) (a){--}(c) $\chi_{aaaa}^{\textrm{S}}(\boldq,0)$ for several $U$ 
with insets showing the static spin susceptibility, 
$\chi^{\textrm{S}}(\boldq,0)=\sum_{a,b}\chi^{\textrm{S}}_{aabb}(\boldq,0)$, 
and (d) $U$ or (e) $J_{\textrm{H}}/U$ dependence of mass enhancement factor, 
$z_{a}(\boldk)^{-1}=
1-\frac{\partial \Sigma_{aa}^{(\textrm{R})}(\boldk,\omega)}
{\partial \omega}|_{\omega \rightarrow 0}$. }
\vspace{-15pt}
\label{fig:Fig1}
\end{figure*}

The parameters in $\epsilon_{ab}(\boldk)$ are chosen so as to reproduce 
the electronic structure of Sr$_{2}$RuO$_{4}$ 
obtained in local-density approximation (LDA)~\cite{LDA}: 
I set $(t_{1},t_{2},t_{3},t_{4},t^{\prime},\Delta_{t_{2g}})=(0.675,0.09,0.45,0.18,0.03,0.13)$ 
and choose $\mu$ so that the total occupation number is four. 
In this choice, 
the total bandwidth is about $4$, 
being twice as large as the experimentally estimated value of $U$~\cite{X-ray10Dq}, 
and the occupation numbers of the $d_{xz/yz}$ and $d_{xy}$ orbitals 
are $n_{xz/yz}=1.38$ and $n_{xy}=1.25$. 
The inconsistency of $n_{xz/yz}$ and $n_{xy}$ with the experimental values~\cite{dHvA-x2} 
($n_{xz/yz}=n_{xy}=1.33$) arises from the quantitative difference that 
the Fermi surface (FS) of the $d_{xy}$ orbital in the LDA~\cite{LDA} 
is closer to the inner sheet in $k_{x}=k_{y}$ line. 
 
The interaction term is treated 
by fluctuation-exchange (FLEX) approximation~\cite{FLEX,multi-FLEX} 
that bubble and ladder diagrams only for electron-hole scattering processes are considered. 
This is suitable for describing 
electronic properties for moderately strong interaction at low $T$ 
since 
this is a perturbation theory beyond MFAs and 
can treat spatial correlation appropriately~\cite{FLEX}.  
By using the procedure~\cite{multi-FLEX} for a paramagnetic phase 
and taking $64^{2}$ meshes of the Brillouin zone and $2048$ Matsubara frequencies, 
I solve the self-consistent equations by iteration 
until the relative error of the self-energy is less than $10^{-4}$. 

The magnetic property and electronic structure of Sr$_{2}$RuO$_{4}$
can be well described in the FLEX approximation. 
First, 
enhancing the spin susceptibility at 
$\boldQ_{\textrm{IC-AF}}=(\frac{21}{32}\pi, \frac{21}{32}\pi)\approx (0.66\pi,0.66\pi)$ 
[Figs. \ref{fig:Fig1}(b) and \ref{fig:Fig1}(c)] 
agrees with the experiment in Ref. ~\onlinecite{Neutron-x2}; 
in contrast to MFAs~\cite{LDA,RPA}, 
its main orbital comes from the $d_{xy}$ orbital. 
This enhancement arises from the combination of 
the self-energy of electrons beyond MFAs 
and orbital-cooperative spin fluctuation: 
the self-energy causes merging of 
the nesting vectors for the $d_{xz/yz}$ and $d_{xy}$ orbitals 
around $\boldQ_{\textrm{IC-AF}}$ due to the FS deformation for the $d_{xy}$ orbital 
and mode-mode coupling for spin fluctuations [Figs. 
\ref{fig:Fig1}(a){--}\ref{fig:Fig1}(c)]; 
this merging leads to enhancing the nondiagonal term of spin fluctuation 
at $\boldQ_{\textrm{IC-AF}}$ between these orbitals; 
this and diagonal terms cause 
the orbital-cooperative enhancement of spin fluctuation at $\boldQ_{\textrm{IC-AF}}$. 
Second, 
the larger mass enhancement~\cite{dHvA-x2} of the $d_{xy}$ orbital 
than that of the $d_{xz/yz}$ orbital is naturally reproduced 
due to the stronger (nonlocal) spin fluctuation of the $d_{xy}$ orbital [Figs. \ref{fig:Fig1}(d) 
and \ref{fig:Fig1}(e)]. 
The agreement with experiment is better than 
that in dynamical-mean-field theory (DMFT)~\cite{Haule-DMFT}. 
Third, 
the values of $n_{xz/yz}$ and $n_{xy}$ are improved in comparison to the LDA values~\cite{LDA}; 
e.g., at $(T, U, J_{\textrm{H}})=(0.006, 1.8, 0.3)$, these are $(n_{xz/yz},n_{xy})=(1.36,1.28)$. 
This improvement is similar to that of the DFMT~\cite{Haule-DMFT}. 

\begin{figure*}[tb]
\includegraphics[width=160mm]{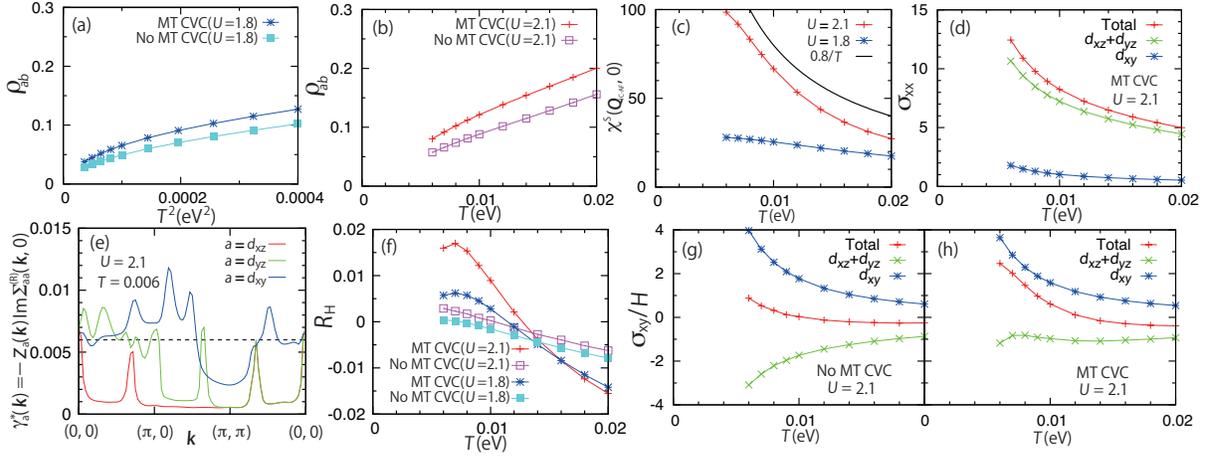}
\vspace{-11pt}
\caption{(Color online) 
(a) $\rho_{ab}$ against $T^{2}$ at $U=1.8$, 
(b) $\rho_{ab}$ against $T$ at $U=2.1$, 
(c) $\chi^{\textrm{S}}(\boldQ_{\textrm{IC-AF}},0)$ against $T$ at $U=2.1$ and $1.8$, 
(d) $\sigma_{xx}$ and orbital components with the MT CVC against $T$ at $U=2.1$, 
(e) the QP damping against $\boldk$ at $(T,U)=(0.006,2.1)$, 
(f) $R_{\textrm{H}}$ against $T$ at $U=2.1$ and $1.8$, 
and $\sigma_{xy}/H$ and orbital components 
(g) without or (h) with the MT CVC against $T$ at $U=2.1$. 
The dashed line in panel (e) corresponds to $T=0.006$.}
\vspace{-10pt}
\label{fig:Fig2}
\end{figure*}

Then, 
I derive $\rho_{ab}$ and $R_{\textrm{H}}$ in the weak-field limit by using the Kubo formulas 
and considering only the most divergent terms~\cite{FL-cond} 
with respect to the QP lifetime~\cite{future-paper}. 
This treatment is correct in the FL and 
remains reasonable in the metallic phases where a perturbation theory works~\cite{FL-basis}.  
In this treatment, 
$\rho_{ab}=\sigma_{xx}^{-1}$ and 
$R_{\textrm{H}}=\sigma_{xy}/H\sigma_{xx}^{2}$ ($\sigma_{yy}=\sigma_{xx}$ is used) 
are determined by 
\begin{align}
\sigma_{xx}=&\ 
\dfrac{2}{N}
\textstyle\sum\limits_{\boldk}
\sum\limits_{\{a\}=1}^{3}
\int^{\infty}_{-\infty}\dfrac{d\epsilon}{2\pi}
\Bigl(-\dfrac{\partial f(\epsilon)}{\partial \epsilon}\Bigr)
\Lambda_{x;ba}^{(0)}(k)\Lambda_{x;dc}(k)\notag\\
&\times G_{ad}^{(\textrm{R})}(k)G_{cb}^{(\textrm{A})}(k),\label{eq:sigmaxx-approx}
\end{align}
and
\begin{align}
\dfrac{\sigma_{xy}}{H}
=&\ \dfrac{1}{N}
\textstyle\sum\limits_{\boldk}
\sum\limits_{\{a\}=1}^{3}
\int^{\infty}_{-\infty}\frac{d\epsilon}{2\pi} 
\Bigl(-\dfrac{\partial f(\epsilon)}{\partial \epsilon}\Bigr)
\Bigl[
\Lambda_{x;ba}(k)
\dfrac{\overleftrightarrow{\partial}}{\partial k_{y}}
\Lambda_{y;dc}(k)
\Bigr]
\notag\\
&\times 
\textrm{Im}
\Bigl[
G_{ad}^{(\textrm{R})}(k)
\dfrac{\overleftrightarrow{\partial}}{\partial k_{x}}
G_{cb}^{(\textrm{A})}(k)
\Bigr].\label{eq:sigmaxy-approx}
\end{align}
Here I use $\sum_{\{a\}}\equiv \sum_{a,b,c,d}$, $k\equiv (\boldk,\epsilon)$, 
and $[
g(x)\frac{\overleftrightarrow{\partial}}{\partial x}h(x)]\equiv
g(x)\frac{\partial h(x)}{\partial x}-\frac{\partial g(x)}{\partial x}h(x)$, 
$G_{ab}^{(\textrm{R}\ \textrm{or}\ \textrm{A})}(k)$ is retarded or advanced Green's function, 
$f(\epsilon)$ is Fermi function, 
$\Lambda_{\nu;ab}^{(0)}(k)$ is renormalized group velocity, 
\begin{align}
\Lambda_{\nu;ab}^{(0)}(k)=\dfrac{\partial \epsilon_{ab}(\boldk)}{\partial k_{\nu}}
+\dfrac{\partial \textrm{Re}\Sigma_{ab}^{(\textrm{R})}(k)}{\partial k_{\nu}},\label{eq:Lambd0}
\end{align} 
where $\Sigma_{ab}^{(\textrm{R})}(k)$ is the retarded self-energy, 
and $\Lambda_{\nu;dc}(k)$ is renormalized current, 
\begin{align}
\Lambda_{\nu;dc}(k)=\Lambda_{\nu;dc}^{(0)}(k)+\Delta \Lambda_{\nu;dc}^{(\textrm{CVC})}(k),\label{eq:Lambd}
\end{align}
with 
\begin{align}
\Delta \Lambda_{\nu;dc}^{(\textrm{CVC})}(k)=&
\dfrac{1}{N}
\textstyle\sum\limits_{\boldk^{\prime}}
\textstyle\sum\limits_{\{A\}=1}^{3}
\int^{\infty}_{-\infty}\frac{d\epsilon^{\prime}}{4\pi i}
\mathcal{J}_{dcCD}^{(0)}(k,k^{\prime})
G_{CA}^{(\textrm{R})}(k^{\prime})\notag\\
&\times G_{BD}^{(\textrm{A})}(k^{\prime})\Lambda_{\nu;AB}(k^{\prime}),\label{eq:CVC}
\end{align} 
where $\mathcal{J}_{dcCD}^{(0)}(k,k^{\prime})$ 
is electron-hole four-point vertex function being irreducible with respect to 
a pair of the retarded and advanced Green's functions. 
$\Delta \Lambda_{\nu;dc}^{(\textrm{CVC})}(k)$ is vital 
to satisfy conservation laws~\cite{Yamada-resistivity} 
since it plays the similar role to the backflow correction. 

To calculate $\Lambda_{\nu;dc}(k)$, 
I use MT four-point vertex function in the FLEX approximation,
\begin{align}  
\mathcal{J}_{abcd}^{(0)}(k,k^{\prime})
=
2i\Bigl(\coth \tfrac{\epsilon-\epsilon^{\prime}}{2T}
+\tanh \tfrac{\epsilon^{\prime}}{2T}\Bigr)\textrm{Im}V_{acbd}^{(\textrm{R})}(k-k^{\prime}),\label{eq:MTCVC} 
\end{align}
where $V_{acbd}^{(\textrm{R})}(q)$ is retarded effective interaction 
in this approximation~\cite{multi-FLEX}. 
This treatment will be sufficient for a qualitative discussion 
since 
the neglected terms~\cite{AL}, being of higher order with respect to the QP damping, 
are much smaller than the MT term in the single-orbital case~\cite{Kontani-CVC} 
and the similar result will hold in the present case. 
Thus, I believe 
the FLEX approximation including the MT CVC is suitable to analyze 
the transport properties of the metallic phases not far away from the AF QCP. 

We turn to results of $\rho_{ab}$ and $R_{\textrm{H}}$. 
Several quantities as a function of $\epsilon$ 
are calculated by the Pad\'{e} approximation~\cite{Pade-approx} 
using the data for the lowest four Matsubara frequencies. 
The $\epsilon$ and $\epsilon^{\prime}$ integrations are done 
by discretizing the interval $0.0025$ and replacing the upper and lower values 
by $0.1$ and $-0.1$.  
$\Lambda_{\nu;dc}(k)$ is calculated by iteration until its relative error is less than $10^{-4}$; 
the singularity of the principal integral for 
the term containing $\coth \tfrac{\epsilon-\epsilon^{\prime}}{2T}$ 
is removed by the $\epsilon^{\prime}$ derivatives of its numerator and denominator 
by using $\textrm{Im}V_{dCcD}^{(\textrm{R})}(\boldq,0)=0$.  

We first compare $\rho_{ab}$ at $U=1.8$ and $2.1$ in Figs. \ref{fig:Fig2}(a) and \ref{fig:Fig2}(b); 
hereafter, I consider $U=2.1$ ($U=1.8$) case near (away from) the AF QCP 
since $\chi^{\textrm{S}}(\boldQ_{\textrm{IC-AF}},0)$ shows 
the Curie-Weiss-like (Pauli paramagnetic) $T$ dependence [Fig. \ref{fig:Fig2}(c)]. 
$\rho_{ab}$ with or without the MT CVC is roughly proportional 
to $T^{2}$ at $U=1.8$ and to $T$ at $U=2.1$. 
Thus, 
the power of the $T$ dependence of $\rho_{ab}$ is determined by the self-energy 
and becomes one near the AF QCP. 

\begin{figure*}[tb]
{\includegraphics[width=124mm]{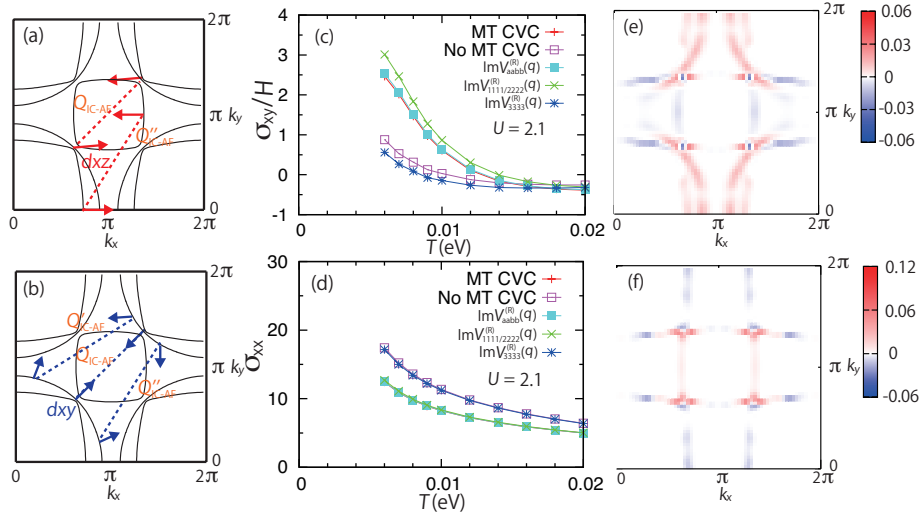}}
\vspace{-12pt}
\caption{(Color online) 
Schematic pictures of the currents of (a) the $d_{xz}$ and (b) the $d_{xy}$ orbital 
connected by the MT CVC, 
(c) $\sigma_{xy}/H$ and (d) $\sigma_{xx}$ against $T$ at $U=2.1$ for several special cases 
whose data are obtained by using part of $\textrm{Im}V_{acbd}^{(\textrm{R})}(q)$ as the CVC, 
(e) $\sigma_{xy}(\boldk)/H$ with the MT CVC against $\boldk$ at $(T,U)=(0.006,2.1)$, and 
(f) the difference between the $d_{xz}+d_{yz}$ components 
of $\sigma_{xy}(\boldk)/H$ with and without the MT CVC against $\boldk$ at $(T,U)=(0.006,2.1)$.}
\vspace{-12pt}
\label{fig:Fig3}
\end{figure*}

To reveal the role of each $t_{2g}$ orbital in $\rho_{ab}$, 
orbital components of $\sigma_{xx}$ with the MT CVC at $U=2.1$ 
are shown in Fig. \ref{fig:Fig2}(d); 
the component of the $d_{xz}$ and $d_{yz}$ orbitals or the $d_{xy}$ orbital 
is calculated from the equation that  
$\sum_{\{a\}=1}^{3}$ in Eq. (\ref{eq:sigmaxx-approx}) is replaced 
by $\sum_{\{a\}=1}^{2}$ or $\sum_{\{a\}=3}$, respectively. 
The main contribution to $\sigma_{xx}$ ($\sigma_{yy}$) comes from 
the $d_{xz}$ ($d_{yz}$) orbital in contrast to that of the spin fluctuation. 
This result arises from 
the smaller QP damping and 
larger renormalized group velocity of the $d_{xz/yz}$ orbital 
than those of the $d_{xy}$ orbital. 
Note that the similar results are obtained at $U=1.8$ (not shown). 

In addition, 
the QP damping of the $d_{xz}$ orbital around $\boldk= \boldQ_{\textrm{IC-AF}}$ 
becomes a hot spot at $U=2.1$, 
although that around $\boldk=(\frac{23}{32}\pi,0)\approx (0.72\pi,0)$ 
remains a cold spot [Fig. \ref{fig:Fig2}(e)]. 
[At the cold (hot) spot, 
the QP damping is (is not) much smaller than temperature considered.] 
Thus, 
the origin of the $T$-linear $\rho_{ab}$ at $U=2.1$ 
is the hot-spot structure of the QP damping of the $d_{xz/yz}$ orbital 
around $\boldk= \boldQ_{\textrm{IC-AF}}$. 
I emphasize that 
this $T$-linear $\rho_{ab}$ is not due to a breakdown of perturbation theory. 

We next compare $R_{\textrm{H}}$ at $U=2.1$ and $1.8$ in Fig. \ref{fig:Fig2}(f). 
There are two main and four secondary results. 
The main results are, first, that 
the peak of $R_{\textrm{H}}$ at $T=0.007$ is induced by the MT CVC 
at $U=2.1$ and $1.8$; second, that 
the Curie-Weiss-like $T$ dependence of $R_{\textrm{H}}$ is absent at $U=2.1$, 
although $\chi^{\textrm{S}}(\boldQ_{\textrm{IC-AF}},0)$ shows the Curie-Weiss-like behavior. 
The first secondary result is that 
the difference between $R_{\textrm{H}}$ without the MT CVC at $U=2.1$ and $1.8$ is small, 
although the QP dampings are different. 
This arises from  
the small effects of the QP damping 
since its effects on $\sigma_{xy}/H$ and $\sigma_{xx}^{2}$ are nearly canceled out. 
The second is that 
the values of these $R_{\textrm{H}}$ are nearly zero. 
The third is that at $U=2.1$ and $1.8$, 
the MT CVC causes the positive enhancement of $R_{\textrm{H}}$ 
in the range of $0.006\leq T\leq 0.012$ 
and the negative enhancement of $R_{\textrm{H}}$ in the range of $0.014\leq T \leq 0.02$. 
The fourth is that 
the positive enhancement at $U=2.1$ is larger than that at $U=1.8$, 
while the negative enhancement at $U=2.1$ is of the same order of magnitude as that at $U=1.8$. 

To understand the two main and last three secondary results, 
I present orbital components of $\sigma_{xy}/H$, 
calculated in a similar way to $\sigma_{xx}$, 
without or with the MT CVC at $U=2.1$ 
in Fig. \ref{fig:Fig2}(g) or \ref{fig:Fig2}(h); 
the following results (i){--}(iv) remain qualitatively the same at $U=1.8$ (not shown). 
(i) The sign of the component of the $d_{xz}$ and $d_{yz}$ orbitals is minus, 
and that of the $d_{xy}$ orbital is plus. 
(ii) The components of the $d_{xz}$ and $d_{yz}$ orbitals and the $d_{xy}$ orbital 
without the MT CVC are nearly the same in magnitude. 
Thus, 
the nearly zero $R_{\textrm{H}}$ without the MT CVC arises from 
the comparable and opposite-sign components of these orbitals. 
(iii) The magnitude decrease for the $d_{xz/yz}$ orbital 
due to the MT CVC is larger than that for the $d_{xy}$ orbital 
in the range of $0.006\leq T\leq 0.012$, 
while the magnitude decrease for these $t_{2g}$ orbitals are very small 
in the higher-$T$ region. 
Combining this with the effect of the MT CVC on $\sigma_{xx}$, 
we find that 
the positive enhancement of $R_{\textrm{H}}$ in the low-$T$ region arises from 
the combination of the decrease of $\sigma_{xx}^{2}$ 
and positive enhancement of $\sigma_{xy}/H$ due to the MT CVC, 
and that 
the negative enhancement of $R_{\textrm{H}}$ in the high-$T$ region 
arises from the combination of 
the decrease of $\sigma_{xx}^{2}$ due to the MT CVC and 
the negative sign of $\sigma_{xy}/H$ with the MT CVC. 
In addition, 
the larger positive enhancement of $R_{\textrm{H}}$ at $U=2.1$ than at $U=1.8$ 
arises from the larger reduction of $\sigma_{xx}^{2}$ due to the MT CVC, 
and the small magnitude difference between the negative enhancement at $U=2.1$ and $1.8$ 
arises from the small effects of the MT CVC on $\sigma_{xx}$ and $\sigma_{xy}/H$ at high $T$. 
(iv) The component of the $d_{xz/yz}$ orbital with the MT CVC 
shows the similar peak to that of $R_{\textrm{H}}$, 
although such peak does not appear in the total component. 
[Note that no peak in $\sigma_{xy}/H$ does not contradict with the peak in $R_{\textrm{H}}$ 
since $R_{\textrm{H}}$ is $(\sigma_{xy}/H) \times \sigma_{xx}^{-2}$.] 
This result implies 
the peak and the absence of the Curie-Weiss-like enhancement of $R_{\textrm{H}}$ 
are related to the orbital dependence of the MT CVC. 

Then, 
I analyze how the MT CVC affects the current of each orbital. 
Combining Eqs. (\ref{eq:Lambd0}){--}(\ref{eq:MTCVC}) 
with the facts in the present model that 
$\Lambda_{\nu;aa}^{(0)}(k)$ are much larger than $\Lambda_{\nu;ab(\neq a)}^{(0)}(k)$ 
due to the larger intraorbital hopping integrals 
and that the dominant terms of $\textrm{Im}V_{acbd}^{(\textrm{R})}(q)$ 
are $\textrm{Im}V_{aabb}^{(\textrm{R})}(q)$ 
due to stronger spin fluctuation than other fluctuations, 
we find 
the dominant effects of the MT CVC in the present model 
are the connections between the intraorbital terms of the currents 
at $\boldk$ and $\boldk^{\prime}$ near the Fermi level. 
In particular, 
since the main terms of $\textrm{Im}V_{aabb}^{(\textrm{R})}(q)$ 
are the low-$\omega$ terms at $\boldq=\boldQ_{\textrm{IC-AF}}$ and 
the secondary are the low-$\omega$ terms at 
$\boldq=\boldQ_{\textrm{IC-AF}}^{\prime}=(\pi,\frac{21}{32}\pi)\approx (\pi, 0.66\pi)$ 
or $\boldQ_{\textrm{IC-AF}}^{\prime\prime}=(\frac{21}{32}\pi,\pi)\approx (0.66\pi, \pi)$ (not shown), 
we see from Figs. \ref{fig:Fig3}(a) and \ref{fig:Fig3}(b), 
first, that 
the main effects are the magnitude decreases of the currents 
of the $d_{xz/yz}$ and $d_{xy}$ orbitals at $\boldk= \boldQ_{\textrm{IC-AF}}$ 
near the Fermi level, 
arising from the low-$\omega$ terms of 
$\textrm{Im}V_{1111/2222}^{(\textrm{R})}(\boldQ_{\textrm{IC-AF}},\omega)$ 
and $\textrm{Im}V_{3333}^{(\textrm{R})}(\boldQ_{\textrm{IC-AF}},\omega)$, respectively; 
second, that the secondary effects are 
the magnitude decrease of the current of the $d_{xz}$ [$d_{yz}$] orbital 
at $\boldk=(\frac{23}{32}\pi, 0)\approx (0.72\pi, 0)$ 
[$(0, \frac{23}{32}\pi)\approx (0,0.72\pi)$] 
due to the low-$\omega$ terms of 
$\textrm{Im}V_{1111}^{(\textrm{R})}(\boldQ_{\textrm{IC-AF}}^{\prime\prime},\omega)$ 
[$\textrm{Im}V_{2222}^{(\textrm{R})}(\boldQ_{\textrm{IC-AF}}^{\prime},\omega)$] 
and the angle changes of the current of the $d_{xy}$ orbital 
at $\boldk=(\frac{7}{8}\pi, 0)\approx (0.88\pi, 0)$ and 
$(0, \frac{7}{8}\pi)\approx (0, 0.88\pi)$ 
due to the low-$\omega$ terms of $\textrm{Im}V_{3333}^{(\textrm{R})}(\boldq,\omega)$ 
at $\boldq= \boldQ_{\textrm{IC-AF}}^{\prime\prime}$ and $\boldQ_{\textrm{IC-AF}}^{\prime}$, 
respectively. 
In addition to these main and secondary effects, 
the MT CVCs arising from the low-$\omega$ terms of $\textrm{Im}V_{aaaa}^{(\textrm{R})}(q)$ 
whose $\boldq$ slightly differs from $\boldQ_{\textrm{IC-AF}}$ or $\boldQ_{\textrm{IC-AF}}^{\prime}$ or 
$\boldQ_{\textrm{IC-AF}}^{\prime\prime}$ 
cause the angle changes of the corresponding currents near the Fermi level. 

Among these effects, 
the most important effect on $R_{\textrm{H}}$ arises from 
the magnitude decrease of the current of the $d_{xz/yz}$ orbital around $\boldQ_{\textrm{IC-AF}}$ 
near the Fermi level. 
One of the facts is that  
the $T$ dependence of $\sigma_{xy}/H$ and $\sigma_{xx}$ with the MT CVC 
are almost reproduced by using the MT CVC 
arising from $\textrm{Im}V_{1111/2222}^{(\textrm{R})}(q)$ [Figs. \ref{fig:Fig3}(c) 
and \ref{fig:Fig3}(d)]. 
This orbital dependence arises mainly from the smaller QP damping of the $d_{xz/yz}$ orbital 
than that of the $d_{xy}$ orbital 
since the kernel of the MT CVC for the $d_{xz/yz}$ [$d_{xy}$] orbital 
is inversely proportional to the QP damping of the $d_{xz/yz}$ [$d_{xy}$] orbital 
and proportional to $\textrm{Im}V_{1111/2222}^{(\textrm{R})}(q)$ [$\textrm{Im}V_{3333}^{(\textrm{R})}(q)$]. 
The other is that 
the most drastic effect of the MT CVC on $\sigma_{xy}/H$ is 
the positive enhancement of the $d_{xz}+d_{yz}$ component of $\sigma_{xy}(\boldk)/H$ 
around $\boldk=\boldQ_{\textrm{IC-AF}}$, 
while the secondary is the negative enhancement of 
that around $\boldk= (\frac{11}{16}\pi, \frac{\pi}{4})\approx (0.69\pi, 0.25\pi)$ 
and $(\frac{\pi}{4}, \frac{11}{16}\pi)\approx (0.25\pi, 0.69\pi)$ 
[Figs. \ref{fig:Fig3}(e) and \ref{fig:Fig3}(f)]. 

From those results of the effects of the MT CVC, 
we find  
the peak of $R_{\textrm{H}}$ with the MT CVC 
arises from the peak of the $d_{xz}+d_{yz}$ component of $\sigma_{xy}/H$ 
as a result of the competition between 
the positive enhancement around $\boldk= \boldQ_{\textrm{IC-AF}}$ 
and negative enhancement around 
$\boldk\approx (0.69\pi, 0.25\pi)$ and $(0.25\pi, 0.69\pi)$ 
due to the MT CVC arising from spin fluctuations of the $d_{xz/yz}$ orbital. 

In addition, 
combining the orbital dependence of the MT CVC 
with the equations of the dependence of $\sigma_{xx}$ 
and $\sigma_{xy}/H$ on the leading order of the angle change, 
$\Delta \varphi_{ab}(k)=\varphi_{ab}(k)-\varphi_{ab}^{(0)}(k)$, 
which are, respectively,   
\begin{align}
\hspace{-5pt}
|\Lambda_{ba}^{(0)}(k)|\cos \varphi_{ba}^{(0)}(k)
|\Lambda_{dc}(k)|\cos \varphi_{dc}^{(0)}(k)[1-\tfrac{\Delta \varphi_{dc}(k)^{2}}{2}]
\label{eq:sigmaxx-leadingPhi}
\end{align} 
and 
\begin{align}
&|\Lambda_{ba}(k)|\cos \varphi_{ba}^{(0)}(k)
|\Lambda_{dc}(k)|\cos \varphi_{dc}^{(0)}(k)
\tfrac{\partial \varphi_{dc}(k)}{\partial k_{y}}\notag\\
+&
|\Lambda_{ba}(k)|\sin \varphi_{ba}^{(0)}(k)
\tfrac{\partial \varphi_{ba}(k)}{\partial k_{y}}
|\Lambda_{dc}(k)|\sin \varphi_{dc}^{(0)}(k), \label{eq:sigmaxy-leadingPhi}
\end{align}
we find 
the absence of the Curie-Weiss-like enhancement of $R_{\textrm{H}}$ near the AF QCP
arises from the absence of the angle change of the current due to the main term of the MT CVC. 
Note that 
although the Curie-Weiss-like $T$-dependent spin fluctuation leads to 
the Curie-Weiss-like $T$ dependence of the magnitude and angle changes of the current 
through $\textrm{Im}V_{aabb}^{(\textrm{R})}(q)$ in the MT CVC, 
the effects of its $T$ dependence of the magnitude change 
on $\sigma_{xy}/H$ and $\sigma_{xx}^{2}$ are nearly canceled out, 
while the Curie-Weiss like $T$ dependence of the angle change appearing 
in $\sigma_{xy}/H$ causes the Curie-Weiss-like enhancement of $R_{\textrm{H}}$. 

Before comparison with experiment, 
I remark on main similarities and differences between the present case 
and the single-orbital case~\cite{Kontani-CVC} 
and propose the realization of the similar transport properties in other systems. 

For $\rho_{ab}$, 
the similarity is the $T$-linear dependence near the AF QCP, 
and the difference is 
the difference between the main orbitals for $\rho_{ab}$ and spin fluctuation. 
This difference arises from the facts that 
$\sigma_{xx}$ is inversely proportional to the QP damping within the leading order, 
and that the strong spin fluctuation enhances the QP damping. 
Since these facts hold in metallic phases of other multiorbital systems, 
this orbital-dependent transport is realized in other systems. 
It should be noted that 
due to this difference in the main orbital, 
the criticality of $\rho_{ab}$ (i.e., the power of its $T$ dependence) 
is not always connected with 
the criticality of fluctuation (i.e., the kind of the QCP) in multiorbital systems. 
In the present case, 
these criticalities become the same due to orbital-cooperative enhancement 
of spin fluctuation at $\boldQ_{\textrm{H}}$.

For $R_{\textrm{H}}$, 
the similarity is the considerable effects of the MT CVC on its low-$T$ values, 
and the differences are the absence of the Curie-Weiss-like $T$ dependence 
and the peak without the peak of the $T$ dependence of the spin susceptibility. 
Since the former difference is related to the $\boldk$ dependence of the main term 
of the MT CVC, as explained, 
this finding gives another ubiquitous mechanism for the $T$ dependence of $R_{\textrm{H}}$ 
near an AF QCP: 
the Curie-Weiss-like $T$-dependent spin fluctuation, characterizing the AF QCP, 
does not cause the Curie-Weiss-like $T$ dependence of $R_{\textrm{H}}$ 
if the directions of the currents connected by the MT CVC arising from 
this spin fluctuation are antiparallel. 
This will be realized in some single-orbital or multiorbital systems near an AF QCP. 
In addition, 
the peak of $R_{\textrm{H}}$ will be realized 
in some metallic phases satisfying four conditions (e.g., some transition metal oxides 
and organic conductors): 
electron correlation is strong; 
quasi-$1$D orbitals form the conducting bands; 
there are opposite-sign components of $\sigma_{xy}(\boldk)/H$ of these orbitals; 
there are at least two nesting vectors for these orbitals, 
each of which affects each component of $\sigma_{xy}(\boldk)/H$ 
through the MT CVC arising from the corresponding spin fluctuation. 
These conditions are necessary for the competition between the opposite-sign enhancement of 
these opposite-sign components of $\sigma_{xy}(\boldk)/H$ of the quasi-$1$D orbitals 
due to the MT CVC arising from spin fluctuations of these orbitals. 

Finally, we compare the results with experiment. 
The results with the MT CVC at $U=1.8$ 
reproduce experimental results~\cite{resistivity-x2,Hall-x2} of Sr$_{2}$RuO$_{4}$, 
the $T$-square $\rho_{ab}$, monotonic increase of $R_{\textrm{H}}$ in $0.007\leq T \leq 0.02$, 
crossing of $R_{\textrm{H}}$ over zero, and peak of $R_{\textrm{H}}$ at $T\sim 0.007$. 
(Although those~\cite{resistivity-x2,Hall-x2} are reproduced 
in relaxation-time approximation~\cite{Hall-theory-x2}, neglecting all the CVCs, 
by choosing some parameters of the QP damping, 
I do not use any such parameters.) 
Since the small quantitative difference in 
the value of $T$ where $R_{\textrm{H}}$ crosses over zero 
(which is $0.014$ in an experiment~\cite{Hall-x2}) 
is related to the difference in the occupation numbers, 
an analysis by the model having the same occupation numbers is a future work.  
Then, the results with the MT CVC at $U=2.1$ 
can explain the $T$-linear $\rho_{ab}$~\cite{Ti214-nFL} in Sr$_{2}$Ru$_{0.075}$Ti$_{0.025}$O$_{4}$. 
Since the measurement of $R_{\textrm{H}}$ in Sr$_{2}$Ru$_{0.075}$Ti$_{0.025}$O$_{4}$ 
has been restricted to a low-$T$ value~\cite{Ti214-RH}, 
the $T$ dependence of $R_{\textrm{H}}$ obtained near the AF QCP 
can be tested in further measurement 
if the main effect of Ti substitution can be assumed to make the system near the AF QCP 
compared with Sr$_{2}$RuO$_{4}$. 

In summary, 
I have studied several electronic properties of the ruthenates near and away from the AF QCP 
in the FLEX approximation including the MT CVC. 
I have found, first, that 
the enhancement~\cite{Neutron-x2} of spin fluctuation at $\boldQ_{\textrm{IC-AF}}$ arises from 
the combination of the self-energy of electrons beyond MFAs 
and orbital-cooperative spin fluctuation; 
second, that 
the larger mass enhancement~\cite{dHvA-x2} of the $d_{xy}$ orbital arises from 
the stronger spatial correlation of that orbital; 
third, that 
the nonmonotonic $T$ dependence of $R_{\textrm{H}}$~\cite{Hall-x2} arises from 
the competition between 
the opposite-sign enhancement of $\sigma_{xy}(\boldk)/H$ of the $d_{xz}$ and $d_{yz}$ orbitals 
around $\boldk=\boldQ_{\textrm{IC-AF}}$ 
and $\boldk\approx (0.69\pi, 0.25\pi)$ and $(0.25\pi, 0.69\pi)$ 
due to the MT CVCs arising from spin fluctuations of these orbitals. 
Also, 
I have explained that 
the $T$-linear $\rho_{ab}$ of Sr$_{2}$Ru$_{0.075}$Ti$_{0.025}$O$_{4}$~\cite{Ti214-nFL} 
can be understood as the hot-spot structure of the QP damping of the $d_{xz/yz}$ orbital 
around $\boldk= \boldQ_{\textrm{IC-AF}}$. 
I have proposed, first, that 
the $T$ dependence of $R_{\textrm{H}}$ near the AF QCP can be experimentally tested 
in Sr$_{2}$Ru$_{0.075}$Ti$_{0.025}$O$_{4}$ 
if the main effect of Ti substitution can be assumed to tune the system 
to the vicinity of the AF QCP; 
second, that 
multiorbital systems in a metallic phase 
show the inplane transport whose main orbital differs from that for spin fluctuation; 
third, that 
some strongly correlated electron systems having quasi-$1$D orbitals show 
the peak of $R_{\textrm{H}}$ at low $T$ 
without the peak of the $T$ dependence of the spin susceptibility; 
fourth, that 
the absence of the Curie-Weiss-like enhancement of $R_{\textrm{H}}$ near an AF QCP 
is realized in some single-orbital or multiorbital systems 
where the angle change of the current due to the main term of the MT CVC is absent.

\begin{acknowledgments}
I thank K. Ueda, H. Tsunetsugu, M. Imada, A. Fujimori, and S. Nakatsuji 
for some meaningful questions 
and useful comments. 
I also thank T. Nomura for a good question about the spin fluctuation of Sr$_{2}$RuO$_{4}$. 
All the numerical calculations were performed 
at the Supercomputer Center in the Institute for Solid State Physics 
at the University of Tokyo. 
\end{acknowledgments}


\begin{thebibliography}{99}
\bibitem{Moriya-review}
T. Moriya, J. Magn. Magn. Mater. \textbf{14}, 1 (1979).

\bibitem{cuprate-exp}
S. W. Tozer, A. W. Kleinsasser, T. Penney, D. Kaiser, and F. Holtzberg, 
Phys. Rev. Lett. \textbf{59}, 1768 (1987); 
T. Penney, S. von Moln\'ar, D. Kaiser, F. Holtzberg, and A. W. Kleinsasser, 
Phys. Rev. B \textbf{38}, 2918 (1988).
 
\bibitem{Ti214-nFL} 
N. Kikugawa and Y. Maeno, 
Phys. Rev. Lett. \textbf{89}, 117001 (2002).

\bibitem{CSRO}
S. Nakatsuji and Y. Maeno, 
Phys. Rev. Lett. \textbf{84}, 2666 (2000); 
L. M. Galvin, R. S. Perry, A. W. Tyler, A. P. Mackenzie, 
S. Nakatsuji, and Y. Maeno,  
Phys. Rev. B \textbf{63}, 161102(R) (2001).

\bibitem{resistivity-x2}
N. E. Hussey, A. P. Mackenzie, J. R. Cooper, 
Y. Maeno, S. Nishizaki, and T. Fujita, 
Phys. Rev. B \textbf{57}, 5505 (1998).

\bibitem{Hall-x2}
A. P. Mackenzie, N. E. Hussey, A. J. Diver, 
S. R. Julian, Y. Maeno, S. Nishizaki, and T. Fujita,  
Phys. Rev. B \textbf{54}, 7425 (1996).

\bibitem{Kontani-CVC} 
H. Kontani, K. Kanki, and K. Ueda, 
Phys. Rev. B \textbf{59}, 14723 (1999); 
Y. Yanase, J. Phys. Soc. Jpn. \textbf{71}, 278 (2002).

\bibitem{MT}
K. Maki,  
Prog. Theor. Phys. \textbf{40}, 193 (1968); 
R. S. Thompson, 
Phys. Rev. B \textbf{1}, 327 (1970).

\bibitem{LDA} 
T. Oguchi, 
Phys. Rev. B \textbf{51}, 1385 (1995); 
I. I. Mazin and D. J. Singh, 
Phys. Rev. Lett. \textbf{79}, 733 (1997). 

\bibitem{X-ray10Dq}
H.-J. Noh, S.-J. Oh, B.-G. Park, J.-H. Park, J.-Y. Kim, H.-D. Kim, 
T. Mizokawa, L. H. Tjeng, H.-J. Lin, C. T. Chen, S. Schuppler, 
S. Nakatsuji, H. Fukazawa, and Y. Maeno, 
Phys. Rev. B \textbf{72}, 052411 (2005).

\bibitem{dHvA-x2}
A. P. Mackenzie, S. R. Julian, A. J. Diver, G. J. McMullan, M. P. Ray,
G. G. Lonzarich, Y. Maeno, S. Nishizaki, and T. Fujita, 
Phys. Rev. Lett. \textbf{76}, 3786 (1996).

\bibitem{FLEX}
N. E. Bickers, D. J. Scalapino, and S. R. White, 
Phys. Rev. Lett. \textbf{62}, 961 (1989); 
N. E. Bickers and S. R. White, 
Phys. Rev. B \textbf{43}, 8044 (1991).

\bibitem{multi-FLEX}
T. Takimoto, T. Hotta, and K. Ueda, 
Phys. Rev. B \textbf{69}, 104504 (2004); 
H. Ikeda, R. Arita, and J. Kune\ifmmode \check{s}\else \v{s}\fi{}, 
ibid. \textbf{81}, 054502 (2010).

\bibitem{Neutron-x2}
Y. Sidis, M. Braden, P. Bourges, B. Hennion, S. NishiZaki, 
Y. Maeno, and Y. Mori, 
Phys. Rev. Lett. \textbf{83}, 3320 (1999).

\bibitem{RPA}
T. Nomura and K. Yamada, 
J. Phys. Soc. Jpn. \textbf{69}, 1856 (2000); 
T. Takimoto, 
Phys. Rev. B \textbf{62}, R14641 (2000); 
N. Arakawa and M. Ogata, 
ibid. \textbf{87}, 195110 (2013). 

\bibitem{Haule-DMFT}
J. Mravlje, M. Aichhorn, T. Miyake, K. Haule, G. Kotliar, and A. Georges, 
Phys. Rev. Lett. \textbf{106}, 096401 (2011).

\bibitem{FL-cond} 
G. M. $\acute{\textrm{E}}$liashberg, 
Sov. Phys. JETP \textbf{14}, 886 (1962); 
H. Kohno and K. Yamada, 
Prog. Theor. Phys. \textbf{80}, 623 (1988). 

\bibitem{future-paper}
Their derivations are going to be published elsewhere. 

\bibitem{FL-basis}
Applicability of a perturbation theory differs from 
that of the FL. 
The FL becomes an approximate eigenstate 
if the QP damping is much smaller than temperature considered; 
otherwise, the FL does not. 
Even in the latter case, 
a perturbation theory works 
if perturbation expansion has a good convergence or becomes an asymptotic expansion. 
For example, see 
P. Nozi$\grave{\textrm{e}}$res, 
\textit{Theory of Interacting Fermi Systems} 
(Addison-Wesley, MA, 1997); 
K. Yamada, \textit{Electron Correlation in Metals} 
(Cambridge University Press, Cambridge, 2004); 
M. J. Rice, 
Phys. Rev. \textbf{159}, 153 (1967); 
H. Ikeda, S. Shinkai, and K. Yamada, 
J. Phys. Soc. Jpn. \textbf{77}, 064707 (2008).

\bibitem{Yamada-resistivity}
K. Yamada and Y. Yosida,
Prog. Theor. Phys. \textbf{76}, 621 (1986).

\bibitem{AL}
L. G. Aslamasov and A. I. Larkin, 
Sov. Phys. Solid State \textbf{10}, 875 (1968).

\bibitem{Pade-approx}
H. J. Vildberg and J. W. Serene, 
J. Low Temp. Phys. \textbf{29}, 179 (1977).

\bibitem{Hall-theory-x2}
C. Noce and M. Cuoco, 
Phys. Rev. B \textbf{62}, 9884 (2000).

\bibitem{Ti214-RH}
N. Kikugawa, A. P. Mackenzie, C. Bergemann, and Y. Maeno,  
Phys. Rev. B \textbf{70}, 174501 (2004).


\end{thebibliography}

\end{document}